\documentclass{pasa}%

\title[A NEW POTENTIAL FORMULA APPLICABLE TO FLATTENED SYSTEMS]{A NEW POTENTIAL FORMULA APPLICABLE TO FLATTENED SYSTEMS}
\author[S. Ninkovi\' c]{S. Ninkovi\' c$^1$\\
\affil{$^1$Astronomical Observatory, Volgina 7, 11060
Belgrade 74, Serbia}%
}%
\jid{PASA}
\doi{10.1017/pas.\the\year.xxx}
\jyear{\the\year}

\begin{document}%
\begin{abstract}
A new formula for the gravitational potential   of flattened systems
is proposed. It is a modification of the Miyamoto-Nagai   potential and
should be applied to very flattened 
systems, exponential discs as a typical example. 
The resulting rotation curve agrees sufficiently
well with that obtained by using special functions and 
the total masses   remain the same.
The functions contained in the new term
can improve the agreement for the rotation 
curve and also reduce the effect of
negative density values which appear
  off the midplane.  

\end{abstract}
\maketitle%
\section{INTRODUCTION }
\label{sec:intro}

The gravitational potential proposed by Miyamoto and
Nagai (1975) (MN) is well known. Its applications to
models of the Milky Way and other disc galaxies are 
numerous (e. g.  Allen Santillan 1991, Ninkovi\' c 1992, Dinescu et al. 1999,
MacMillan \& Binney 2011). The observational evidence is in
favour of exponential discs characterised by the 
dependence of the circular speed on the 
distance to the centre (rotation curve) as given in
Freeman's paper (1970). As to the fitting of the 
circular-speed dependence resulting from the
Miyamoto-Nagai formula to 
Freeman's curve, there are some difficulties 
indicated in an earlier paper of the present author 
(Ninkovi\' c 2003a). In the present paper a new formula 
for the gravitational potential is proposed. It appears 
as a generalisation of MN and it contains 
the formula analysed in the earlier paper (Ninkovi\' c 2003a)
as a special case. The density generating this potential as
function of the coordinates is also studied.

\section{THE POTENTIAL FORMULA}

  The gravitational potential in stellar systems
(subsystems), such as, for instance bulges and discs of spiral galaxies,
  is often represented by a variant of the Green formula (e. g. 
Cuddeford 1993). This would mean that in general the gravitational potential
$\Phi $ should have the following form


\begin{equation}\label{eq1}
{\Phi}={G {\cal M} \over {\cal R}} \ \ ,
\end{equation}

\noindent{where $G $ and ${\cal M} $ are constants, 
whereas ${\cal R} $ is a   function}. 
The first constant ($G $) is the universal gravitation 
constant, the second one (${\cal M} $) is the total mass of 
the system (subsystem) appearing as the source of the
gravitational field. ${\cal R} $ is a function 
of  the generalised coordinates and time. It must satisfy the
following conditions: to have dimension of length and 
to be approximately equal to the distance from the centre
of the source system $r $ at the points lying very far 
from the system centre.   

The potential of a flattened 
stellar system, as here of interest, is assumed to depend on
two arguments: $R $, distance to the axis of symmetry $z $, and 
$\vert z \vert $, distance to the midplane. The function
proposed by Miyamoto and Nagai (1975) is a well-known example

\begin{equation}\label{eq2}
{\cal R}_{MN}=\bigg[R^2 + \Big(a+ \sqrt{z^2+b^2}\Big)^2\bigg]^{1/2} \ \ .
\end{equation}

\noindent{In this function $a $ and $b $ are   constants and for  
significantly flattened systems satisfy $b/a < 1 $. When
the function ${\cal R}_{MN} $ is substituted in equation (1), one obtains 
the Miyamoto-Nagai potential formula. The present author proposes here
a potential formula which contains a new term   ${\cal R}_N $, as follows}

\begin{equation}\label{eq3}
\Phi={G{\cal M} \over {{\cal R}_{MN}-{\cal R}_N}} \ \ .
\end{equation}

\noindent{In a general case the new term ${\cal R}_N $ is also a 
function of the same two variables, $R $ and $\vert z \vert $ and 
is everywhere positive. Since ${\cal R}_{MN} $ satisfies the
condition of being approximately equal to the distance $r $ at
the points lying very far from the system centre, the new term
at such points must be negligibly small compared to ${\cal R}_{MN} $.
In this way the difference ${\cal R}_{MN}-{{\cal R}_N} $ will satisfy
the condition of being approximately equal to the distance from the
centre $r $ at very distant points.}

\section{THE ROTATION CURVE}

The   surface density of the disc of a spiral galaxy, bearing in mind the observational constraints,
is usually assumed to be exponential (often referred to as Freeman law), for instance its luminosity 
or mass density $\rho $ obeys the following formula (e. g. Deg \& Widrow 2013, their equation (13))

\begin{equation}\label{eq4}
\rho(R,z)=\rho(0) exp\bigg(-{R \over R_d}\bigg) sech^2\bigg({z \over z_d}\bigg) \ \ ,
\end{equation}

\noindent{where $R_d $ and $z_d $ are constants. As easily seen, the
surface density following from this expression will depend on $R $ as 
a simple exponential function with $R_d $ as the scale length.} 

Unfortunately, equation (4) yields no analytical solutions for the potential via the 
Poisson equation. This is important because with the known potential 
it is possible to obtain the circular speed $u_c $ by using the well-known relation 

\begin{equation}\label{eq5}
u_c=\sqrt{-R{\partial \Phi \over {\partial R}}} \ \ , \ \ z=0  \ \ .
\end{equation}

\noindent{The plot   of circular speed versus distance $R $ according to Freeman's solution 
(Fig. 1) is characterised by existence of radius at which the circular speed is the same
for the Keplerian and exponential disc cases of the same mass. 
As a consequence,   at infinity the circular speed for the exponential disc 
approaches the Keplerian curve from "above"  , i. e. from higher values of $u_c $. This is not the case with the 
potential formula of Miyamoto and Nagai. Substituting equation (2) in equation (1)
and applying (5) one obtains the dependence of the circular speed on $R $ 
  corresponding to the Miyamoto-Nagai potential. Since $z=0 $, the parameters $a $
and $b $ enter the circular-speed dependence always through their sum $a+b $. In 
other words, the flattening expressed by means of $b/a $, does not affect the 
circular-speed dependence. The maximum circular speed occurs at   about $1.4(a+b) $ and 
the circular-speed value resulting from the Miyamoto-Nagai formula is everywhere
smaller than the value yielded by the Keplerian dependence for the same total mass. 
Due to this a good fit between the circular speed following from the potential
of Miyamoto and Nagai and that corresponding to the exponential disc is achieved
at the cost of different total masses. Usually the total mass in the Miyamoto-Nagai
formula is about 1.5 times as large as that of the exponential disc   (e. g. Ninkovi\' c 1992).}
 
The purpose of introducing the new term ${{\cal R}_N} $ (equation (3)) is to reproduce 
the property of intersection with the Keplerian curve and to have the same total mass as 
for the exponential disc. As shown earlier (Ninkovi\' c 2003a), this is possible even
with a constant substituted for ${{\cal R}_N} $. The constant introduced here will be
$R_d $, the scale length. Then in the fitting   the rotation curve one should determine the ratio $(a+b)/R_d $.
According to the definition assumed in the present paper (equation (1)) the potential cannot have 
negative values, so this ratio must be greater than 1.   We find a best fit value of 2.1. The fit   can be further 
improved. This is done
by generalising the term ${{\cal R}_N} $ in the following way

\begin{equation}\label{eq6}
{\cal R}_N={1 \over 2}R_d\Bigg[\bigg(1+{R^2 \over c_1^2}\bigg)^{\gamma_1}+\bigg(1+{z^2 \over c_2^2}\bigg)^{\gamma_2}\Bigg] \ \ ,
\end{equation}

\noindent{where $\gamma_1<0.5$, $\gamma_2<0.5$ Though it may seem that the number of parameters now tends to be too 
large,   this is not the case in practice. For instance, $R_d $ is quite acceptable to be 
substituted for $c_1 $. The second term in equation (6) (the function of $\vert z \vert $)
is foreseen because of the density   distribution the potential returns (see next section). With regard to all 
equations written above except (4) one obtains the following expression which yields the circular speed}

\begin{eqnarray}\label{eq7}
u_c&=&\sqrt{{G{\cal M}R \over 
\Bigg[\sqrt{R^2 + (a+b)^2}-{\frac{1}{2}R_d\bigg(1+\Big(1+{R^2 \over R_d^2}\Big)^{\gamma_1}\bigg)}\Bigg]^2}}\cdot \nonumber  \\
&&\cdot \sqrt{{{R \over \sqrt{R^2 + (a+b)^2}}}-{\gamma_1 {R \over R_d} \Bigg(1+{R^2 \over R_d^2}\Bigg)^{\gamma_1-1}}}
\end{eqnarray}

\begin{figure}
\begin{center}
\includegraphics[width=8.5cm]{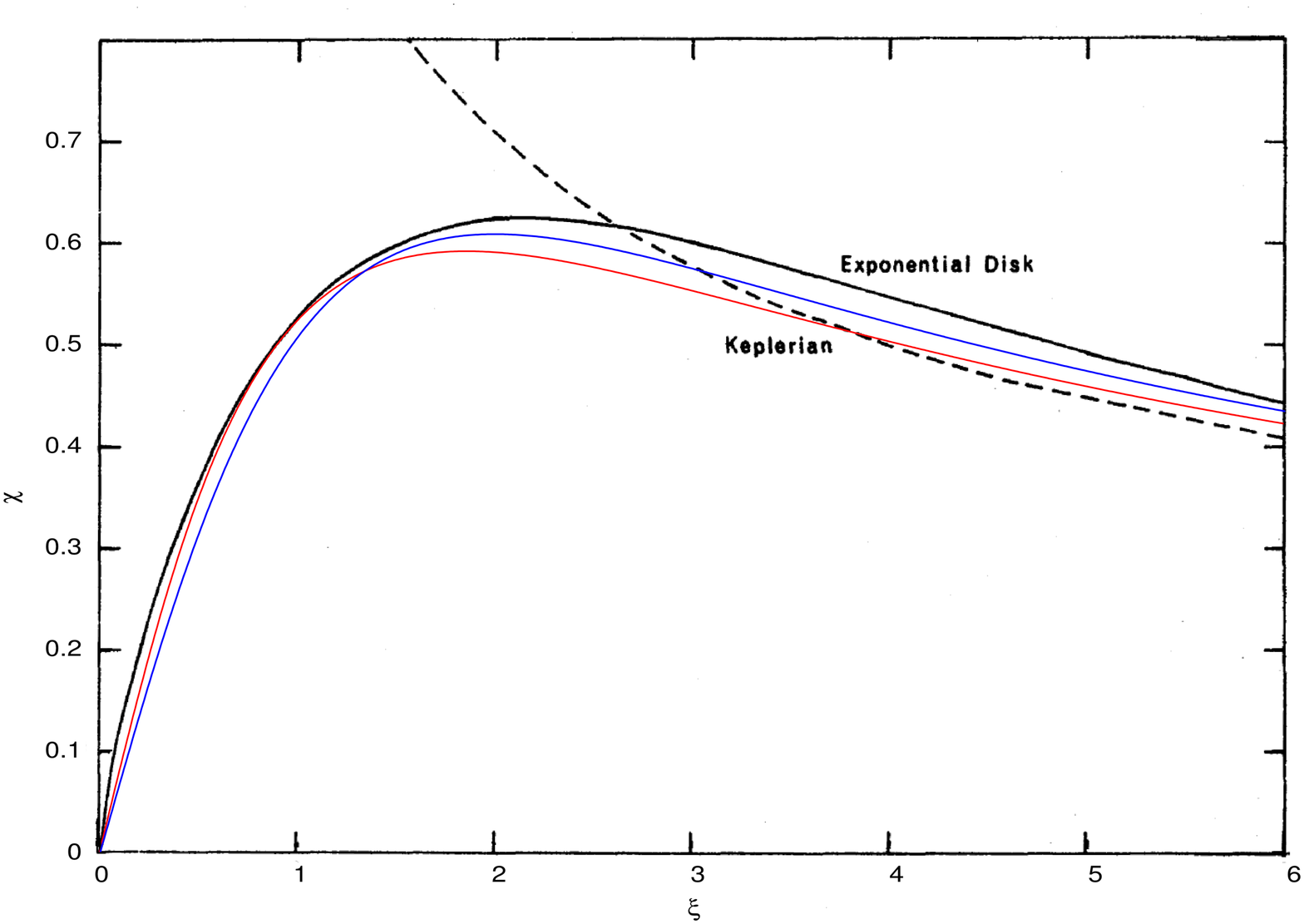}
\caption{Rotation curves from equation (7) as described in the text (blue curve $\gamma_1=-0.1 $ , red curve $\gamma_1=-0.3 $) together with Freeman's (1970) curve; $\xi=R/R_d $, 
$\chi=u_c\bigg/\sqrt{{G{\cal M}\over R_d}} $}
\end{center}
\end{figure}

The curves presented in Fig. 1 follow from equation (7) with $a+b=2.1 R_d $, $\gamma_1=-0.1 $ and $\gamma_1=-0.3 $. They are
presented together with the corresponding curve from Freeman's (1970) paper. The dimensionless 
variables, $\xi $ and $\chi $, are defined as: $\xi=R/R_d $, $\chi=\sqrt{G{\cal M}/R_d} $. The
comparison requires the scale lengths $R_d $ and the total masses ${\cal M} $ to be equal 
in both formalisms.

\section{THE DENSITY}

The density which generates a gravitational 
potential and the potential are related through the Poisson 
equation          

\begin{equation}\label{eq8}
\nabla ^2 \Phi=-4\pi G \rho \ \ .
\end{equation}

\noindent{$\nabla^2 $ is the Laplacian, $\rho $
is the density. The full formula is given in 
the Appendix. 

In the density calculation the constants $c_1 $ and $c_2 $ (equation (6)), as well as 
the ratio $b/a $, must be specified. The fit for the rotation curve has yielded
$a+b=2.1 R_d $, under the condition $b\ll a $ (flattened mass distribution),
so it is easy to conclude that $a $ is approximately $2R_d $. In such conditions the 
first term within the brackets in equation (6) should not differ substantially from 1. As
for the second one, a too small $c_2 $ can contribute to significant values of ${\cal R}_N $,
for instance $c_2=b $, leads to such a high ${\cal R}_N $ that the denominator in equation (3)
becomes negative and, as a consequence, the potential, contrary to the convention assumed here, 
becomes also negative. Thus, it is reasonable to assume $c_1=c_2=R_d $.}

In spite of introducing the new term ${\cal R}_N $ (equations (3) and (6)) negative density values 
cannot be avoided. In the midplane the density is nowhere negative. Its dependence on $R $ in $z=0 $
is a monotonously decreasing function. As for the dependence $\rho(z) $, $R=const $, there are three types 
of profiles: i) with a minimum $\rho_{min}<0 $ and then approaching zero from the 
negative side (Fig. 2); ii) a wavy profile, with one minimum and one maximum and then approaching zero, along $R=0 $ both
extrema can be positive, in general the minimum is negative, the maximum positive (Fig. 3); 
iii) a sound profile, without negative values and monotonously decreasing, but obtained along $R=0 $ only (Fig. 4). 
In order to be more clearly visible the most essential part of the curve is magnified and given in the same figure (Fig. 2 and Fig. 3).

In the simplest case - $\gamma_1=\gamma_2=0 $ (equation (6)) - the density dependence $\rho(z) $, $R=const $,
is always of type i) (Fig. 2). The $z $ value at which the density minimum occurs depends on $R $, the smaller 
$R $ is, the closer is the minimum to the midplane. Finally, when $R $ tends to infinity, the $z $ value also 
tends to infinity, because the density tends to zero. 

When the term ${\cal R}_N $ (equations (3) and (6)) takes a functional form, i. e. both $\gamma_1 $ and $\gamma_2 $ 
are non-zero, the dependence $\rho(z) $, $R=const $, changes the form, towards the types ii) and iii). 
For $\gamma_1 $ some kind of critical value appears at which the dependence $\rho(z) $, $R=0 $, obtains the
form iii) (Fig. 4). It should be $\gamma_1 <0$, the modulus is affected by the ratio $b/a $. For instance,
for the pair of values $b=0.18 $, $a=1.92 $, the critical value is $\gamma_1=-0.3 $, for $b=0.08 $, $a=2.02 $, 
the corresponding is $\gamma_1=-0.5 $, for $b=0.02 $, $a=2.08 $, $\gamma_1=-0.6 $, etc. The influence of 
$\gamma_2 $ is not so strong; it is better to be $\gamma_2 >0$, but increasing $\gamma_2 $ leads to weaker  
decreasing of $\rho $ with increasing $z $. The most important is that increasing the modulus of  
$\gamma_1 $ further does not improve the situation, as the negative density values remain and the wavy 
profiles 
survive (Fig.  3). However, the moduli of both exponents $\gamma_1 $ and $\gamma_2 $ are limited. As for the former, 
the main limitation comes from the fit of the rotation curve, in the case of the latter, since it is supposed
to be positive, the physical limitation ($\gamma_2<0.5 $) becomes essential. The final conclusion is that the best
achievable result is to obtain a monotonously decreasing density along $R=0 $ and wavy density profiles along 
cylinders $R>0 $. For the purpose of giving a more clear explanation the following example is chosen. The values of 
the parameters in equations (3) and (6) are: $a+b=2.1 R_d $ ($b=0.18 R_d $), $c_1=c_2=R_d $, $\gamma_1=-0.3 $, 
$\gamma_2=0.3 $. The rotation curve (Fig. 1 - red curve) and the density profiles (Fig. 3 and Fig. 4) correspond to this example.

\begin{figure}
\begin{center}
\includegraphics[width=8.5cm]{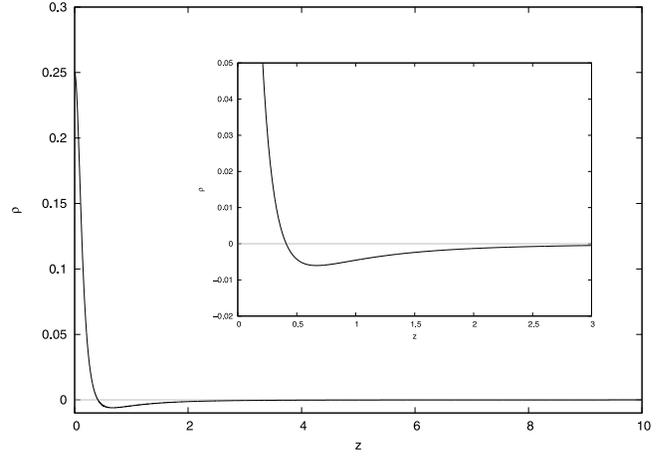}
\caption{Density dependence on height $\vert z \vert $, 
for $R=3 R_d $, $\gamma_1 = \gamma_2 = 0$ (see equation 6); distance unit $R_d$, density unit $\cal{M}$ $R_d^{-3}$}
\end{center}
\end{figure}

\begin{figure}
\begin{center}
\includegraphics[width=8.5cm]{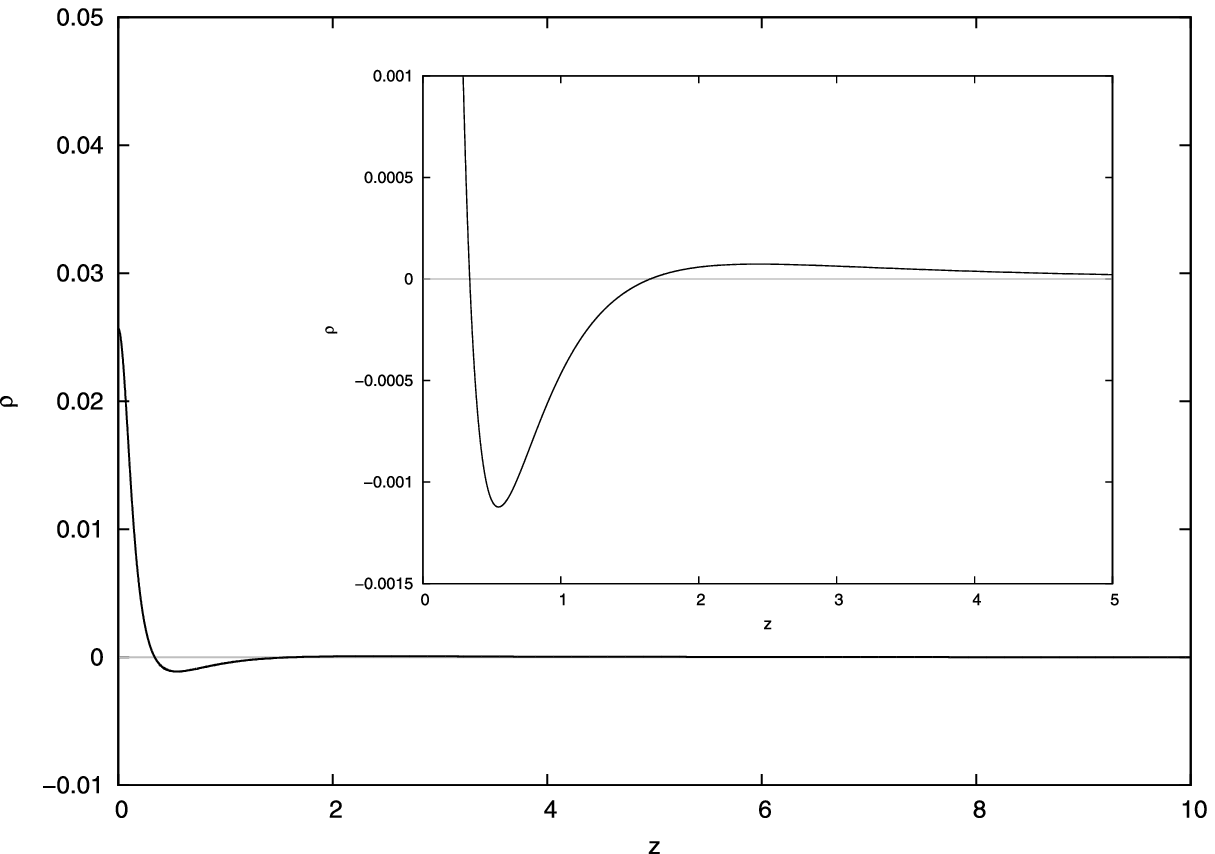}
\caption{Density dependence on height $\vert z \vert $, 
for $R=R_d $, $\gamma_1 = -0.3$ , $\gamma_2 = 0.3$ (see equation 6); distance unit $R_d$, density unit $\cal{M}$ $R_d^{-3}$}
\end{center}
\end{figure}

\begin{figure}
\begin{center}
\includegraphics[width=8.5cm]{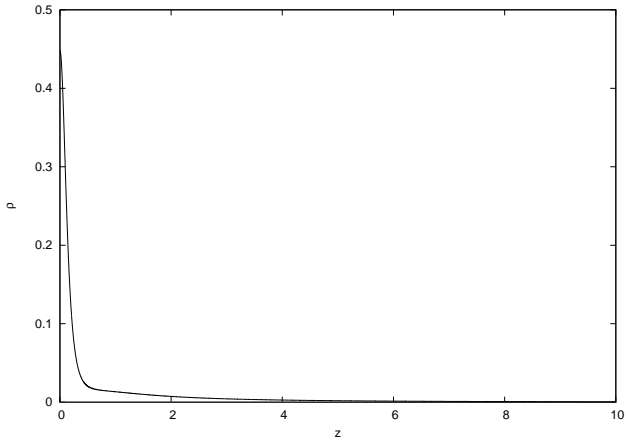}
\caption{Density dependence on height $\vert z \vert $,
for $R=0 $, $\gamma_1 = -0.3$ , $\gamma_2 = 0.3$ (see equation 6); distance unit $R_d$, density unit $\cal{M}$ $R_d^{-3}$}
\end{center}
\end{figure}

In Fig. 2 we show another density comparison, typical for $\gamma_1=\gamma_2=0 $, corresponding to the same  
$a+b $ and the ratio $b/a $. The wave (Fig. 3) can be explained by the rather rapid
density decrease with increasing distance to the midplane. Beyond the centre the density values in the midplane 
are generally low, significantly lower than at the centre, so that at reasonable distances to the midplane 
they are practically zero. Here one obtains the density values by using the Poisson equation (8) which
in the case of axial symmetry means as the algebraic sum of three terms (see Appendix). Clearly, in a numerical 
procedure instead of density values of exactly zero we usually have values of low moduli with both signs. 
In this way the wavy density profiles obtained here can be easily understood as due to numerical precision issues. 

\section{DISCUSSION AND CONCLUSIONS}

The present author proposes an analytical form for the gravitational potential which would 
correspond to the exponential disc, the most luminous subsystem of a spiral galaxy. This is done by
modifying the well-known formula of Miyamoto and Nagai (equations (1)-(3) and (6)). In this modification
there are three essential parameters: the total mass $\cal M $, the exponential scale length $R_d $ and 
the other scale length $b $. The other three parameters - $a $, $\gamma_1 $ and $\gamma_2 $ are auxiliary.
Their role is to improve the fit of the rotation curve particularly, their values are limited. 
Since the system (subsystem) under study is 
flattened, it must satisfy $a\gg b $, and bearing in mind the constraint based on the rotation curve, one 
obtains $a\approx 2R_d $. The exponents $\gamma_1 $ and $\gamma_2 $ cannot exceed 0.5 and $\gamma_1 $ is
constrained to be small and negative, while $\gamma_2 $ lies in $0<\gamma_2<0.5 $. Thus an exponential disc modeled 
in the way as proposed here should be generally characterised by its total mass and two scale lengths. 
      
Though negative density values cannot be avoided completely, the functions in the new term (equation (6)) contribute 
to a significant reducing of this effect. A discussion concerning the analogous spherical-symmetry model can be found
in Ninkovi\' c, 2003b.

\begin{acknowledgements}
This research has been supported by the Serbian Ministry of Education, Science and Technological Development (Project No 176011 "Dynamics and Kinematics of Celestial Bodies and Systems"). The author wants to thank an anonymous referee for valuable suggestions which contributed to a better presentation of the results.
\end{acknowledgements}

\begin{appendix}

\section{APPENDIX}

Equations (1) and (3) from the main text are given again

\begin{equation*}\label{eq9}
\Phi={G{\cal M} \over {{\cal R}}} \ \ ,
\end{equation*}

\begin{equation*}\label{eq10}
{\cal R}={\cal R}_{MN}-{{\cal R}_N} \ \ .
\end{equation*}

{The designations used are: $\Phi $, the gravitational potential, 
$G $ the universal gravitation constant, whereas ${\cal R}_{MN}$ and
${\cal R}_N $ are two functions depending on the arguments $R $
and $z $ (axial symmetry).}  

The Poisson equation ((8) from the main text) relating 
the density and potentail is also rewritten

\begin{equation*}\label{eq11}
\nabla ^2 \Phi=-4\pi G \rho \ \ .
\end{equation*}

{With regard to the axial symmetry the Laplacian has the form}

\begin{equation*}\label{eq12}
\nabla^2\Phi={\partial^2\Phi \over {\partial R^2}}+{1\over R}{\partial \Phi \over {\partial R}}+{\partial^2\Phi \over {\partial z^2}} \ \ .
\end{equation*}

The derivatives will be:
\begin{equation*}\label{eq13}
{\partial \Phi \over {\partial R}} = -{G{\cal M} \over {\cal R}^2}
\Bigg[ {\partial {\cal R}_{MN} \over {\partial R}} -
{\partial {\cal R}_N \over {\partial R}}\Bigg] \ \ ;
\end{equation*}

\begin{eqnarray*}
{\partial^2 \Phi \over {\partial R^2}}&=&
{2G{\cal M} \over {\cal R}^3}
\Bigg[  {\partial{\cal R}_{MN} \over \partial R}-
{\partial{\cal R}_N \over \partial R}\Bigg]^2-\\
&&-{G{\cal M} \over {{\cal R}^2}}
\Bigg[{\partial^2{\cal R}_{MN} \over {\partial R^2}}-
{\partial^2{\cal R}_N \over {\partial R^2}}\Bigg] \ \ ;
\end{eqnarray*}

\begin{eqnarray*}
{\partial^2 \Phi \over {\partial z^2}}&=&
{2G{\cal M} \over {\cal R}^3}
\Bigg[  {\partial{\cal R}_{MN} \over \partial z}-
{\partial{\cal R}_N \over \partial z}\Bigg]^2-\\
&&-{G{\cal M} \over {{\cal R}^2}}
\Bigg[{\partial^2{\cal R}_{MN} \over {\partial z^2}}-
{\partial^2{\cal R}_N \over {\partial z^2}}\Bigg] \ \ ;
\end{eqnarray*}

The first function in the denominator ${\cal R}_{MN}$ and its necessary derivatives are:

\begin{equation*}\label{eq16}
{\cal R}_{MN}=\bigg[R^2 + \Big(a+ \sqrt{z^2+b^2}\Big)^2\bigg]^{1/2} \ \ ;
\end{equation*}

\begin{equation*}\label{eq17}
{\partial{\cal R}_{MN} \over {\partial R}}={R \over{{\cal R}_{MN}}} \ \ ;
\end{equation*}

\begin{equation*}\label{eq18}
{\partial^2{\cal R}_{MN} \over {\partial R}^2}={{{\cal R}_{MN}-{R^2 \over {\cal R}_{MN}}} \over {{\cal R}_{MN}^2}} \ \ ;
\end{equation*}

\begin{equation*}\label{eq19}
q=a+\sqrt{z^2+b^2} \ \ ;
\end{equation*}

\begin{equation*}\label{eq20}
{\partial{\cal R}_{MN} \over \partial z}=
{q \over {{\cal R}_{MN}}} {\partial q \over \partial z} \ \ ;
\end{equation*}

\begin{equation*}\label{eq21}
{\partial^2{\cal R}_{MN} \over {\partial z^2}}=
{q_z{{\cal R}_{MN}-q_z ({q^2 / {\cal R}_{MN}})} 
\over{{\cal R}_{MN}^2}}q_z + {q \over{\cal R}_{MN}}q_{zz} \ \ ;
\end{equation*}

\begin{equation*}\label{eq22}
q_z \equiv {\partial q \over {\partial z}}={z \over {\sqrt{b^2+z^2}}} \ \ , \ \ 
q_{zz}\equiv{\partial^2q \over {\partial z^2}}={b^2 \over{(b^2+z^2)^{3/2}}} \ \ .
\end{equation*}

The second function in the denominator ${\cal R}_N $ and its necessary derivatives are:

\begin{equation*}\label{eq23}
{\cal R}_N={1 \over 2}R_d\Bigg[\bigg(1+{R^2 \over R_d^2}\bigg)^{\gamma_1}+\bigg(1+{z^2 \over R_d^2}\bigg)^{\gamma_2}\Bigg] \ \ ;
\end{equation*}

\begin{equation*}\label{eq24}
{\partial{\cal R}_N \over {\partial R}}=
\gamma_1{R \over{R_d}}\bigg(1+{R^2 \over{R_d^2}}\bigg)^{\gamma_1-1} \ \ ;
\end{equation*}

\begin{eqnarray*}\label{eq25}
{\partial^2{\cal R}_N \over {\partial R^2}}&=&
2\gamma_1(\gamma_1-1){R^2 \over{R_d}^3}\bigg(1+{R^2 \over {R_d^2}}\bigg)^{\gamma_1-2}+\\
&&+{\gamma_1 \over{R_d}}\bigg(1+{R^2 \over {R_d^2}}\bigg)^{\gamma_1-1} \ \ ;
\end{eqnarray*}

\begin{equation*}\label{eq26}
{\partial{\cal R}_N \over {\partial z}}=\gamma_2{z \over{R_d}}\bigg(1+{z^2 \over {R_d^2}}\bigg)^{\gamma_2-1} \ \ ;
\end{equation*}

\begin{eqnarray*}\label{eq27}
{\partial^2{\cal R}_N \over {\partial z^2}}&=&
2\gamma_2(\gamma_2-1){z^2 \over{R_d}^3}\bigg(1+{z^2 \over{R_d^2}}\bigg)^{\gamma_2-2}+\\
&&+{\gamma_2 \over{R_d}}\bigg(1+{z^2 \over{R_d^2}}\bigg)^{\gamma_2-1} \ \ .
\end{eqnarray*}

\end{appendix}


\end{document}